# Light-emitting diode excitation for upconversion microscopy: a quantitative assessment

*Yueying Cao[1,2], Xianlin Zheng[1,3], Simone De Camillis[1,3], Bingyang Shi[1,4], James A. Piper[1,3], Nicolle H. Packer[1,2], Yiqing Lu[1,3,5,*]*

[1]ARC Centre of Excellence for Nanoscale BioPhotonics (CNBP), [2]Department of Molecular Sciences, [3]Department of Physics and Astronomy, [4]Department of Biomedical Sciences, [5]School of Engineering, Macquarie University, Sydney, New South Wales 2109, Australia

*Correspondence should be addressed to Y.L. (yiqing.lu@mq.edu.au)

**Lanthanide-based upconversion nanoparticles (UCNPs) generally require high power laser excitation. Here we report wide-field upconversion microscopy at single-nanoparticle sensitivity using incoherent excitation of a 970-nm light-emitting diode (LED). We show that due to its broad emission spectrum, LED excitation is about 3 times less effective for UCNPs and generates high background compared to laser illumination. To counter this, we use time-gated luminescence detection to eliminate the residual background from the LED source, so that individual UCNPs with high sensitizer ($Yb^{3+}$) doping and inert shell protection become clearly identified under LED excitation at 1.18 W $cm^{-2}$, as confirmed by correlated electron microscopy images. Hydrophilic UCNPs are obtained by polysaccharide coating via a facile ligand exchange protocol to demonstrate imaging of cellular uptake using LED excitation. These results suggest a viable approach to bypassing the limitations associated with high-power lasers when applying UCNPs and upconversion microscopy to life science research.**

The quest for improved fluorescent/luminescent bioprobes has drawn considerable attention to lanthanide-based upconversion nanoparticles (UCNPs), with a series of major breakthroughs achieved over the past 15 years.[1-3] Consisting of inorganic (typically fluoride) nanocrystalline hosts doped with trivalent lanthanide ions, UCNPs are capable of efficiently combining the absorbed energy of multiple near-infrared (NIR) photons to emit photons of shorter wavelengths. Their photostable anti-Stokes luminescence under excitation wavelengths in biological transparent windows has brought new opportunities for life scientists, who have been coping with conventional fluorescent probes that are prone to fast photobleaching, ubiquitous autofluorescence background and limited light penetration in biological tissues. UCNPs have proven to be particularly powerful for applications such as live-cell longitudinal imaging,[4] high-level multiplexing bioassays,[5, 6] optical super-resolution nanoscopy at low power,[7-9] deep-tissue photodynamic therapy and optogenetic manipulation,[10-12] as well as artificial infrared vision.[13]

Despite these advances, UCNPs generally require excitation using lasers that produce output power of several hundred milliwatts to watts, sometimes even tens of watts.[14, 15] This is because most energy transitions of interest (i.e. $4f$-$4f$ transitions of trivalent lanthanide ions) are Laporte forbidden, resulting in small absorption cross-section at defined wavelengths and low radiative relaxation rates.[16] Although the stepwise energy pumping process in UCNPs is much more efficient compared to other multiphoton processes,[17] the quantum yield of UCNPs is nowhere close to that of the normal Stokes emission. A brief estimation indicates that a typical 40-nm UCNP is only 1/80 as bright as a single green fluorescence protein molecule, even though it already incorporates thousands of emitters inside (SI.1). Moreover, except for certain special configurations (such as total internal reflection fluorescence microscopy[18]), implementation of these high power lasers raises concerns over their biomedical applications, as the thermal effect and phototoxicity may already alter some biological/physiological processes, not to mention potential damage to cells and tissues.[19, 20] In addition, retrofitting such lasers (which in most cases are Class 4 lasers of invisible wavelengths) into analytical instruments requires not only expertise in laser optics, but also meticulous laser safety measures for both researchers and laboratories. These issues are currently impeding the broad adoption and translation of UCNPs in biological and biomedical fields.

To circumvent the challenges associated with laser excitation, here we investigated whether single UCNPs can be detected and imaged using light-emitting diode (LED) excitation under common wide-field epifluorescence microscopy. Specifically, we coupled a commercial 970-nm LED to the back illumination port of an inverted microscope, producing maximum excitation intensity of 2.35 W cm$^{-2}$ at the focal plane after a 60× objective. As a first step, we attempted to image commonly-used hexagonal-phase NaYF$_4$ nanoparticles doped with 20% Yb and 1% Tm, which were dropcast (at 2 mg mL$^{-1}$ in 50 µL cyclohexane, followed by evaporation of the solvent) on a glass coverslip. Under illumination at the maximum power of the 970-nm LED, no upconversion luminescence (UCL) was observed (Figure 1a). Next, we tested highly-doped UCNPs wherein all Y$^{3+}$ in the lattice was replaced by Yb$^{3+}$ as in NaYb$_{0.99}$Tm$_{0.01}$F$_4$ that greatly enhances the absorption at the excitation wavelength.[21-24] A portion of this highly-doped UCNP sample was further modified by addition of an inert NaYF$_4$ shell to the nanoparticles (i.e. core-shell configuration) so as to reduce surface quenching.[25, 26] While the uncoated high-Yb UCNPs produced only very weak UCL under the LED excitation (Figure 1b), we were able to observe strong blue emission from the core-shell UCNPs by naked eye, and were even able to record this using a smartphone camera (Figure 1c).

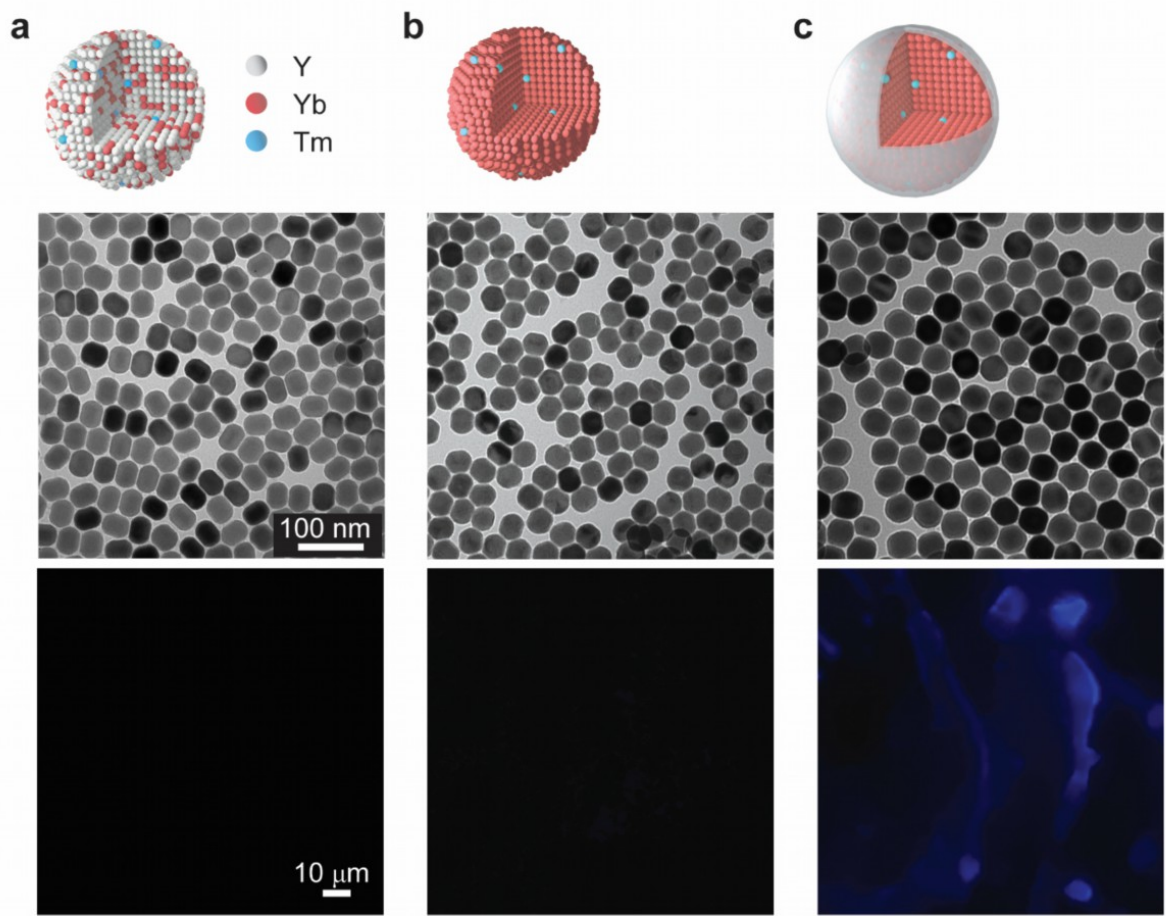

**Figure 1.** TEM and LED-excited UCL images of (a) $NaY_{0.79}Yb_{0.20}Tm_{0.01}F_4$, (b) $NaYb_{0.99}Tm_{0.01}F_4$ and (c) $NaYb_{0.99}Tm_{0.01}F_4@NaYF_4$ UCNPs. Grey, pink and blue dots represent Y, Yb, and Tm ions, respectively. The UCL images were taken by a HUAWEI Mate 10 smartphone in Pro Mode with a blue bandpass filter added before the camera.

The above result confirms that of a previous report[27] that LEDs can excite UCNPs; however, for practical implementation in microscopy, it is critical to make quantitative comparisons of LED excitation to laser excitation. Accordingly, we have measured the absorption spectrum of the core-shell UCNPs ($NaYb_{0.99}Tm_{0.01}F_4@NaYF_4$) and calculated the overlap with the (area normalized) illumination spectra of the 970-nm LED and a 976-nm diode laser, respectively (Figure 2a). Due to the broad spectral distribution of the LED source, the spectral overlap for the LED was 0.36 times that of the laser (SI.2), whose narrow illumination band overlaps nicely with the absorption peak of the UCNPs. That is, for the same excitation intensity, absorption of the 970-nm LED light by the UCNPs is 0.36 times that of the 976-nm laser light. The excitation intensity of 2.35 W cm$^{-2}$ delivered by the LED is therefore equivalent to ~0.8 W cm$^{-2}$ under the laser excitation. This was verified by comparing the UCNP emission spectra, wherein the relative intensities of the emission peaks vary according to the excitation intensity.[22, 28] As shown in Figure 2b, the similarity between the upconversion spectra acquired under the two conditions (LED illumination at 2.35 W cm$^{-2}$ *vs.* laser illumination at 0.79 W cm$^{-2}$) confirmed that LED and laser illumination are indeed interchangeable as long as the former delivers excitation intensity at the focal plane about 3 times higher than that of the latter.

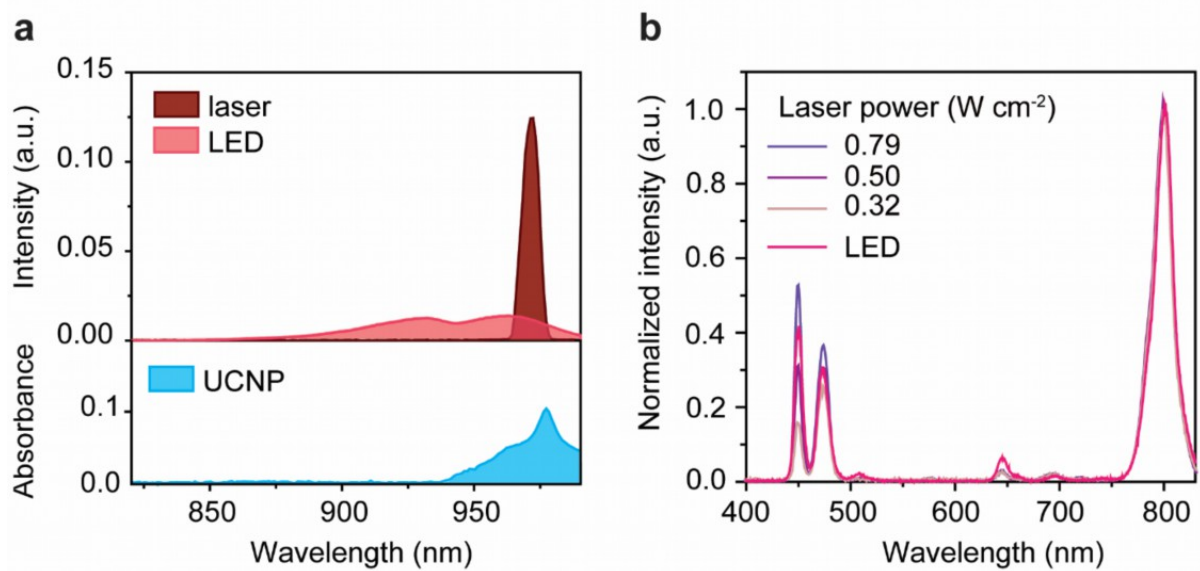

**Figure 2.** Comparison between laser and LED excitation. (a) The excitation spectra of the 976-nm laser and 970-nm LED (top) and the absorption spectrum of NaYb$_{0.995}$F$_4$:Tm$_{0.005}$@NaYF$_4$ (bottom). (b) Emission spectra of NaYb$_{0.995}$F$_4$:Tm$_{0.005}$@NaYF$_4$ under the laser and LED excitation. Each spectrum was normalized to its emission peak at 800 nm.

Next, we investigated the UCNP brightness with respect to the Tm doping concentration (in the core) under this excitation condition. NaYb$_{1-x}$Tm$_x$F$_4$ nanoparticles of similar size (32-42 nm) but with varied Tm doping concentrations ($x$ = 0.25, 0.5, 1, 2 and 4 mol%) were synthesized, all subsequently coated with NaYF$_4$ inert shells of ~3 nm thickness (SI.3). Figure 3a shows the UCL spectra of the core-shell (CS) UCNPs measured in suspension at identical mass concentration in cyclohexane, excited by the 976-nm laser at 0.79 W cm$^{-2}$. It is seen that while these UCNPs appeared blue (Figure 1), the majority of their emission was in the near-infrared (NIR) 800 nm band corresponding to the Tm $^3$H$_4$ → $^3$H$_6$ transition. The UCNPs doped with 0.5% Tm were the brightest in terms of the blue emission between 440 and 490 nm (i.e. 450 and 475 nm bands combined). For the NIR emission, the 1% Tm particles generated the strongest signal, which was 3 times the highest achievable blue emission. By contrast, at the same measurement condition, the core-only high-Yb UCNPs yielded 1.6~76 and 17~58 times weaker emission in the blue and NIR range, respectively (Figure 3b). In addition, the emission of NaY$_{0.79}$Yb$_{0.2}$Tm$_{0.01}$F$_4$ UCNPs was 2 orders-of-magnitude lower than the CS high-Yb UCNPs (SI.4). These results are consistent with previous reports,[22] confirming the significant enhancement of UCL under low excitation intensity by increasing sensitizer concentration and using inert (undoped) shell geometries, resulting in stronger absorption of exciting illumination and reduced surface quenching, respectively.

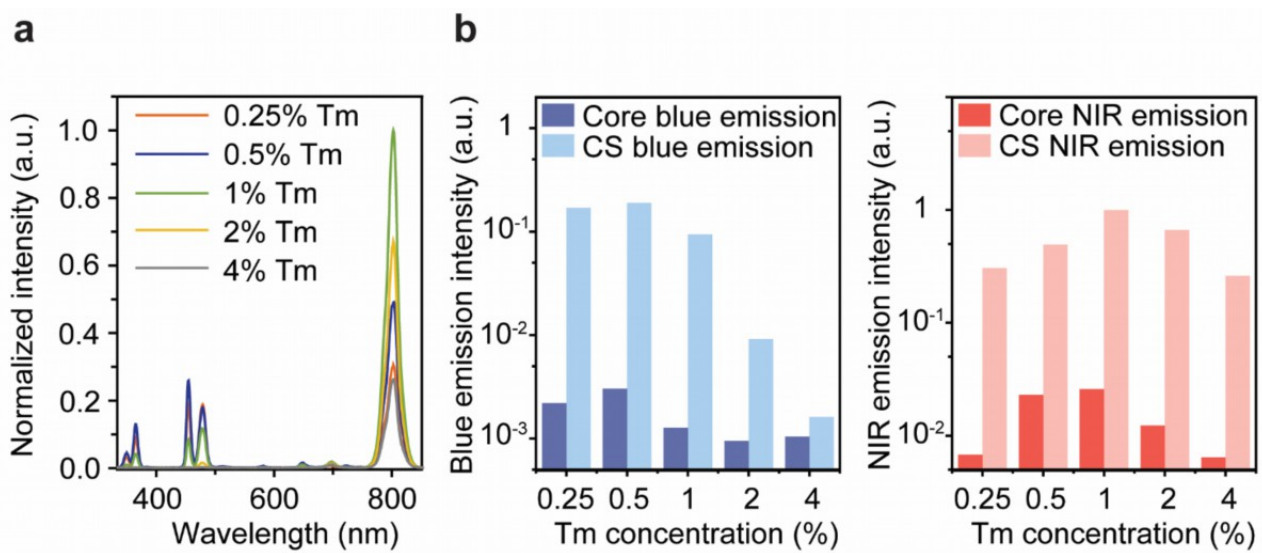

**Figure 3.** Comparison of upconversion emission of UCNPs with varied Yb and Tm doping concentration at equivalent maximum LED excitation intensity. (a) Emission spectra of CS NaYb$_{1-x}$Tm$_x$F$_4$@NaYF$_4$ (x=0.25, 0.5, 1, 2 and 4 mol%) excited by 976-nm laser at 0.79 W cm$^{-2}$. (b) Quantification of blue (440-490 nm) and NIR (775-825 nm) emission of core NaYb$_{1-x}$Tm$_x$F$_4$ and CS NaYb$_{1-x}$Tm$_x$F$_4$@NaYF$_4$ UCNPs excited by 976-nm laser at 0.79 W cm$^{-2}$. The intensities in both figures were normalized to the NIR emission intensity of the 1% Tm CS particles.

To further quantify the brightness, we selected the 1% Tm CS UCNPs and dispersed them on a coverslip to undertake correlative luminescence and scanning electron microscopy (SEM) imaging (SI.5). Taking advantage of the long luminescence lifetime of the UCNPs, we employed the time-gated luminescence (TGL) technique[29] along with an electron-multiplying charge-coupled device (EMCCD) camera to effectively acquire the UCL while eliminating the optical background signal caused by insufficiently filtered LED excitation. As shown in Figure 4a, the enlarged area in the TGL image contained a single UCNP and clusters of 2, 3, and 4 UCNPs, which was confirmed by the corresponding SEM image. Notably, the integrated luminescence signal was proportional to the number of UCNPs in the clusters (Figure 4b), enabling single UCNPs to be distinguished from clusters of 2 or more. The signal-to-noise ratio measured from individual UCNPs was determined to be 1.6 ± 0.1. Based on our experimental conditions and the specification of the EMCCD, the upconversion emission rate from a single NaYb$_{0.99}$Tm$_{0.01}$F$_4$@NaYF$_4$ UCNP with 41 nm core and 3 nm shell was estimated to be 4 photons per second under the LED excitation (SI.6). Note that, since pulsed excitation with 50% duty ratio were used for TGL detection, the effective excitation intensity was reduced by half to 1.18 W cm$^{-2}$. This sets a benchmark for the level of brightness of UCNPs under wide-field microscopy using LED excitation.

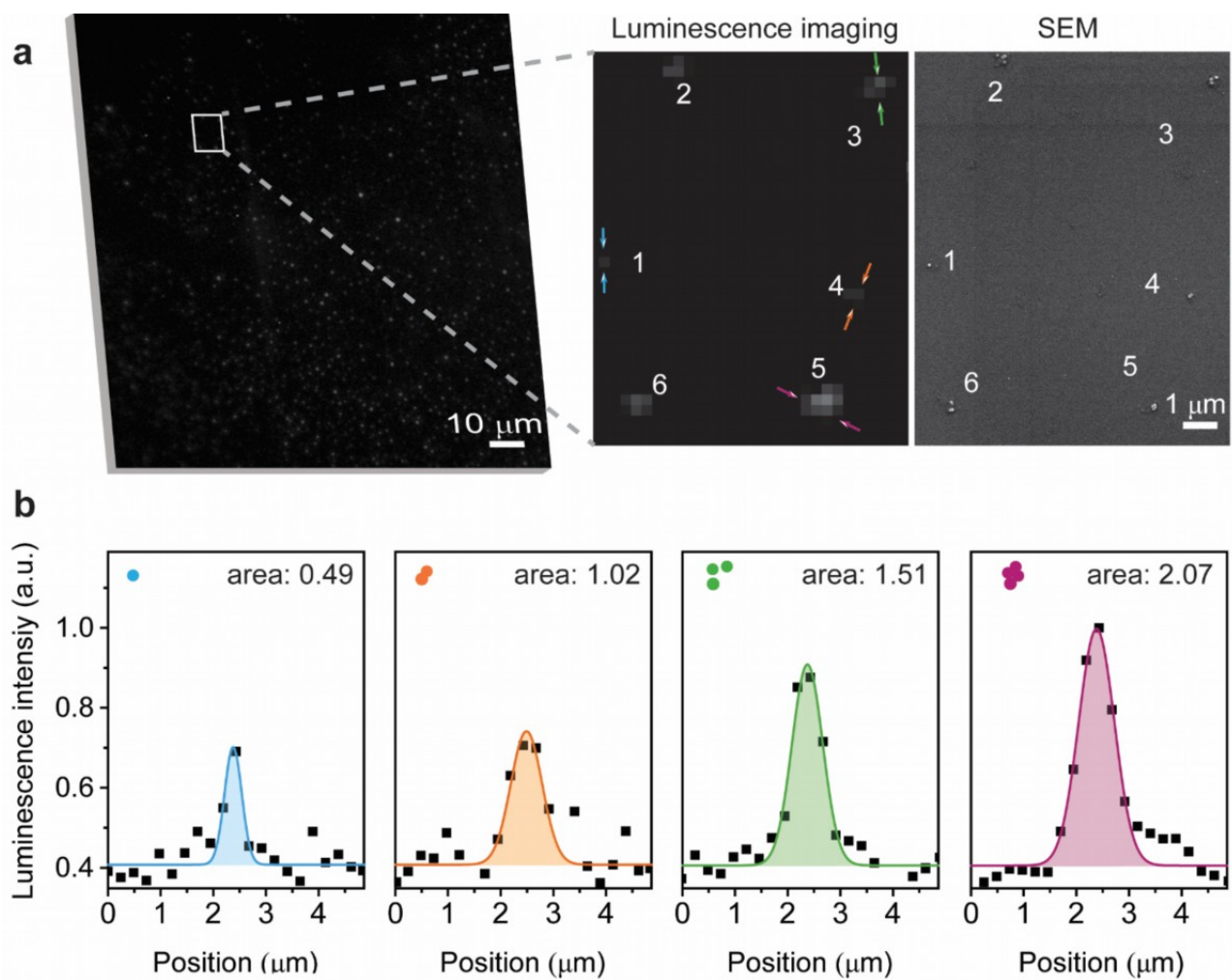

**Figure 4.** Correlative light and electron microscopy images of NaYb$_{0.99}$Tm$_{0.01}$F$_4$@NaYF$_4$ UCNPs. (a) UCL and SEM images of the same sample area. (b) The luminescence intensity along the sections of spot 1, 4, 3 and 5 in (a), which has 1-4 UNCPs, respectively. The TGL image was taken under LED excitation with an exposure time of 30 seconds. The intensity has been normalized by the maximum pixel intensity of the 4-UCNP cluster.

Finally, we applied LED-excited upconversion microscopy to demonstrate cellular imaging. The 1% Tm CS UCNPs were rendered hydrophilic via ligand exchange in a one-phase system that replaces oleic acid on their surface with colominic acid, a polysaccharide rich in carboxyl groups (Figure 5a). Successful conjugation with the negatively charged polysaccharide was verified by dynamic light scattering and zeta-potential measurement, showing that the sugar-coated UCNPs have a hydrodynamic diameter of 59 nm and surface charge of -19 mV (SI.7). The derivatized UCNPs were stable in aqueous solution, with brightness well reserved thanks to the inert shell coating. The hydrophilic UCNPs were then incubated with U87MG glioma-like astrocytes, and showed negligible cytotoxicity to these cells after 24-hour incubation (SI.8). Following actin filament and nuclei staining (by Alexa 488 and DAPI, respectively), the cells were fixed and imaged by TGL microscope under LED excitation, clearly showing the uptake of the UCNPs by the cells (Figure 5b). The same sample slide was doubled checked using a laser scanning confocal microscope equipped with a 976-nm laser, confirming that the UCNPs were indeed inside the cells (SI.9).

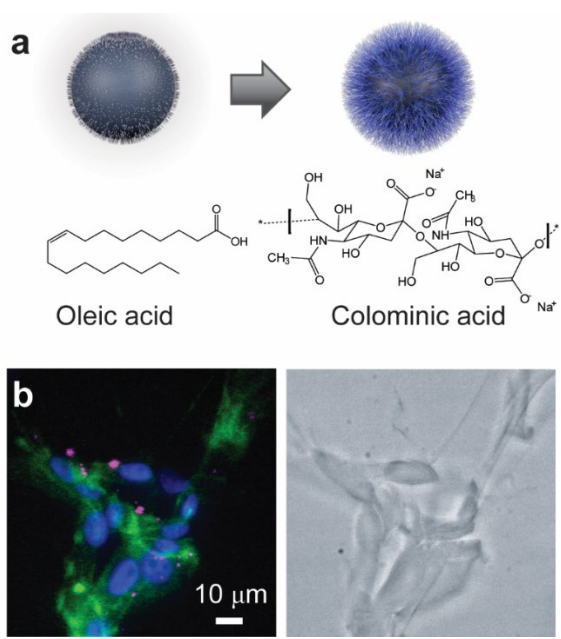

**Figure 5.** Cellular imaging with colominic acid coated UCNPs. (a) Schematics illustrating the surface ligand exchange from oleic acid to colominic acid. (b) *In vitro* imaging (blue-DAPI, green-Alexa 488, magenta-UCNP) of the sugar-coated UCNP incubated with U87MG cells under 1.18 W cm$^{-2}$ LED excitation (left) and the corresponding bright-field image (right). The concentration of UCNPs used for cell co-culture is 5 µg mL$^{-1}$.

This work confirms the viability of LED excitation for upconversion microscopy, achieving sensitivity down to single UCNPs with the help of time-gated luminescence detection. The 970-nm LED excitation of UCNPs was evaluated as 1/3 as effective as for laser excitation, a compromise that is surprisingly small and can be offset by LED sources delivering higher power or by combining multiple LED units together. A difficulty lies in coupling the LED output to the imaging plane of the microscope, due to the large divergence angle associated with its typical incoherent emission. The LED light coupling efficiency was just 0.5% in our current setup, which employs only commercially available components (SI.10). This resulted in relatively weak upconversion luminescence and necessitated long exposure times using the EMCCD camera. However, even small increases in the excitation intensity will lead to very large increases in the luminescence signal, as a result of the nonlinear power dependence of upconversion emission. Special attention to increasing coupling efficiency from the LED may therefore be expected to produce substantial gains in upconversion signal. Additionally, optical super-resolution techniques that are compatible with LED excitation, such as structured illumination microscopy (SIM)[30] and super-resolution radial fluctuations (SRRF)[31, 32], may be further integrated to improve the spatial resolution for LED-excited upconversion microscopy.

From the bioprobe perspective, UCNPs capable of being excited at low irradiance are favored, such as those with the high sensitizer doping and inert core-shell protection employed here. Other approaches to enhancing upconversion luminescence including dye sensitization and plasmonic enhancement[33-38] might be used; however, these may complicate biofunctionalization. If nanoparticle size is not a concern, UCNPs with larger core and thicker shell can be used to provide brighter luminescence, allowing shorter exposure time to detect single nanoparticles. Among the UCNPs we synthesized (despite some variation in nanoparticle size), those with 99% Yb and 1% Tm doping concentrations were shown to produce the brightest upconversion luminescence (mainly in NIR), but the optimum is likely to change depending on the excitation intensity. Hydrophilic and biocompatible UCNPs are readily obtained by ligand exchange with polysaccharides, as well as via other functionalization methods,[39, 40] to enable *in vitro* imaging. The current work provides useful guidance for life scientists seeking to embrace UCNPs and upconversion microscopy for bioimaging without needing to use high-power lasers, which hopefully will significantly improve up-take of these techniques in biomedical research and translation.

**Supporting Information**

Additional materials SI.1-10 include experimental sections, additional figures and notes.

**Acknowledgments**

The authors acknowledge the financial support from the Australian Research Council through the Centre of Excellence for Nanoscale BioPhotonics (CE140100003) and Discovery Early Career Research Award (Y.L., DE170100821). This work was also supported by an International Macquarie University Research Excellence Scholarship (Y.C.) and Macquarie University Research Fellowship (X.Z.). We thank Minakshi Das for helpful discussions and Guoying Wang for assistance with imaging.

**Conflict of Interest**

The authors declare no conflict of interest.